\documentclass[11pt]{article}
\usepackage{graphicx}

\begin{document}

\def\xslash#1{{\rlap{$#1$}/}}
\def \p {\partial}
\def \dd {\psi_{u\bar dg}}
\def \ddp {\psi_{u\bar dgg}}
\def \pq {\psi_{u\bar d\bar uu}}
\def \jpsi {J/\psi}
\def \psip {\psi^\prime}
\def \to {\rightarrow}
\def \lrto{\leftrightarrow} 
\def\bfsig{\mbox{\boldmath$\sigma$}}
\def\DT{\mbox{\boldmath$\Delta_T $}}
\def\xit{\mbox{\boldmath$\xi_\perp $}}
\def \jpsi {J/\psi}
\def\bfej{\mbox{\boldmath$\varepsilon$}}
\def \t {\tilde}
\def\epn {\varepsilon}
\def \up {\uparrow}
\def \dn {\downarrow}
\def \da {\dagger}
\def \pn3 {\phi_{u\bar d g}}

\def \p4n {\phi_{u\bar d gg}}

\def \bx {\bar x}
\def \by {\bar y}

\begin{center}
{\Large\bf  QCD Evolutions of Twist-3 Chirality-Odd Operators  }
\par\vskip20pt
J.P. Ma$^{1,2}$, Q. Wang$^{3}$ and  G.P. Zhang$^{4}$     \\
{\small {\it
$^1$ Institute of Theoretical Physics, Academia Sinica,
P.O. Box 2735,
Beijing 100190, China\\
$^2$ Center for High-Energy Physics, Peking University, Beijing 100871, China  \\
$^3$ Department of Physics and Institute of Theoretical
Physics, Nanjing Normal University, Nanjing, Jiangsu 210097, P.R.China \\
$^4$ School of Physics,  Peking University, Beijing 100871, China
}} \\
\end{center}
\vskip 1cm
\begin{abstract}
We study the scale dependence of twist-3 distributions defined with chirality-odd quark-gluon operators.
To derive the scale dependence we explicitly calculate these distributions of multi-parton states instead of a hadron. 
Taking one-loop corrections into account we obtain the leading evolution kernel in the most general case. 
In some special cases the evolutions are simplified.   
We observe that the obtained kernel in general does not get simplified in the large-$N_c$ limit in contrast to the case 
of those twist-3 distributions defined only with chirality-odd quark operators. In the later, the simplification is significant.  
\vskip 5mm
\noindent
\end{abstract}
\vskip 1cm
\par 
Predictions can be made from QCD with the concept of factorizations for processes with large momentum transfers. 
A typical example is DIS. For unpolarized DIS, the differential cross section  at the leading power of the momentum transfer $Q$ is predicted as a convolution of perturbative coefficient functions with parton distribution functions(PDF's). 
PDF's are defined with twist-2 QCD operators and describe nonperturbative effects of hadrons. 
Although PDF's can not be predicted with perturbative QCD, but they scale dependence, hence the $Q^2$-dependence 
of the differential cross section, can be determined by perturbation theory. 
The dependence is governed by the famous DGLAP equation. In the past,
the predicted scale-dependence or DGLAP equation has played an indispensable role for testing QCD as the correct theory 
of strong interaction.  
\par 
In general, factorized differential cross sections also contain contributions involving hadronic matrix elements 
of higher-twist operators.  Although these contributions are suppressed by inverse powers of $Q$, they contain more informations 
about inner structure of hadrons. Among them, the most interesting 
are those involving twist-3 operators. These contributions are responsible for certain asymmetries in differential cross sections. 
These asymmetries can be measured in experiment and hence provide information about hadronic matrix elements of twist-3 
operators. A well-known example is the study of Single transverse-Spin Asymmetries(SSA). The asymmetries can 
be factorized with the ETQS matrix elements defined with chirality-even quark-gluon 
operators at twist-3\cite{EFTE,QiuSt}.  
The scale dependence of these twist-3 operators have been studied 
in \cite{KQ1,BMP,VoYu,ZhSc,MaWang,KQ2}.  Besides them, there are chirality-odd quark-gluon operators at twist-3. 
In this work, we study the scale
dependence of these operators.

\par 
We consider a spin-1/2 hadron moving in the $z$-direction with the momentum $ P^\mu =(P^+, P^-,0,0)$.   
We will  use the  light-cone coordinate system, in which a
vector $a^\mu$ is expressed as $a^\mu = (a^+, a^-, \vec a_\perp) =
((a^0+a^3)/\sqrt{2}, (a^0-a^3)/\sqrt{2}, a^1, a^2)$ and $a_\perp^2
=(a^1)^2+(a^2)^2$. In this coordinate system we introduce two light-cone vector: $n^\mu =(0,1,0,0)$ and $l^\mu=(1,0,0,0)$. 
There are two distributions can be defined with chirality-odd quark-gluon operators at twist-3. 
They are: 
\begin{eqnarray}
\tilde T_F  (x_1,x_2) &=&   g_s\int\frac{d y_1 dy_2}{4 \pi} e^{-iy_1x_1 P^+ +i y_2 x_2 P^+ } 
   \langle P\vert \bar \psi (y_1 n) \left ( i \gamma_{\perp \mu}\gamma^+ \right )   G^{+\mu}(0)  \psi (y_2 n ) \vert P\rangle, 
\nonumber\\
   \lambda \tilde T_{\Delta}  (x_1,x_2) &=&  g_s\int\frac{d y_1 dy_2}{4\pi} e^{-iy_1x_1 P^+ +i y_2 x_2  P^+} 
   \langle P,\lambda \vert \bar \psi (y_1 n) \left (i\gamma_{\perp\mu} \gamma^+ \gamma_5 \right )   G^{+\mu}(0)  \psi (y_2 n) \vert P,\lambda \rangle,  
\label{tw3}   
\end{eqnarray} 
where the matrix element in the first line is spin-averaged and that in the second line is of a longitudinally polarized 
hadron with the helicity $\lambda=\pm 1/2$. $x_{1,2}$ are momentum fractions. 
The definitions are given in the light-cone gauge $n\cdot G=0$. In other gauges 
gauge links along the direction $n$ should be supplemented to make the definitions gauge invariant. From general principle 
one can show that the two distributions are real and:
\begin{equation} 
\tilde T_F (x_1,x_2) = \tilde T_F(x_2,x_1), \ \ \   \tilde T_\Delta  (x_1,x_2) = -\tilde T_\Delta (x_2,x_1). 
\end{equation} 
Replacing 
the gluon field strength tensor $G^{+\mu}(0)$ in Eq.(\ref{tw3}) with the covariant derivative $D^\mu =\partial^\mu + i g_s G^\mu (0)$, 
one can obtain another two twist-3 distributions. With equation of motion one can relate those two distributions 
to the two distributions defined in Eq.(\ref{tw3}), respectively 
(see, e.g., \cite{JKT, ZYL,BD}). 
\par
The operators in Eq.(\ref{tw3}) contains the operator of the gluon field strength tensor.  There are two chirality-odd twist-3 operators  
which only consist of quark field operators. Correspondingly, one can also define two distributions.   
One is $e(x)$ for unpolarized hadrons, another is $h_L(x)$ for longitudinally polarized hadrons. Again, these two distributions are not independent. One can use operators identities in \cite{VBIF} or equation of motion to show that these two are related with the above two and plus some contributions with local 
operators. The evolution of $e(x)$ and $h_L(x)$ have been studied in \cite{BBKT,KoNi,BD} and the evolution 
equations have been solved in moment space. The evolution of twist-3 quark-gluon operators has been studied in \cite{Beli1} with the emphasis on the solutions of evolution equations. 
 In this work we derive the evolution kernel for the two twist-3 
distributions defined in Eq.(\ref{tw3}) in momentum fraction space with a different method.
\par 
Under renormalization there is no mixing between the two operators in Eq.(\ref{tw3}). The evolution kernels can be conveniently 
expressed with one function by introducing the combinations:
\begin{eqnarray} 
\tilde T_{\pm } (x_1,x_2,\mu) &=&  
  \langle P,\pm \vert {\mathcal O} (x_1,x_2)  \vert P,\pm\rangle
 = \tilde T_F(x_1,x_2,\mu) \pm \tilde T_{\Delta}(x_1,x_2,\mu),
\nonumber\\
  {\mathcal O} (x_1,x_2) &=& g_s \int\frac{dy_1 dy_2}{4\pi} e^{-ix_1 y_1 P^+ + i x_2 y_2 P^+} 
  \psi(y_1n) i\gamma_{\perp\mu} \gamma^+ (1+\gamma_5 ) G^{+\mu}(0) \psi (y_2n).       
\end{eqnarray}
The functions $T_{\pm}(x_1,x_2)$ are nonzero in the region $\vert x_{1,2}\vert \leq 1$ and $\vert x_1-x_2 \vert\leq 1$.  
The scale-dependence can be written in the form:
\begin{equation}
   \frac{\partial}{\partial \ln \mu} \tilde T_{\pm}(x_1,x_2,\mu) 
     =\frac{\alpha_s}{\pi} \int d\xi_1 d\xi_2 {\mathcal F}_{\pm} (x_1,x_2,\xi_1,\xi_2) \tilde T_{\pm}(\xi_1,\xi_2,\mu). 
\label{GEV}      
\end{equation}
The integration region of $\xi_{1,2}$ is fixed by the support of $\tilde T_{\pm}$. It is easy to find that ${\mathcal F}_+$ 
is related to ${\mathcal F}_-$. Here we will take ${\mathcal F}_-$ to give our results.
\par 
The distributions are defined for a given hadron, but the kernel does depend on the hadron. It is completely 
determined by the dynamics of QCD. For large $\mu$ it can be calculated with perturbative QCD. 
Because of this one can use various parton states to calculate the distribution $\tilde T_-$ to find its $\mu$-dependence, 
hence the kernel ${\mathcal F}_-$. For the case with operators of twist-2 one can use single-parton state 
for the purpose. 
But for the two distributions 
defined here, one simply finds null results with single-parton states because the chirality is flipped by the operators. 
Therefore, one has to use multi-parton states to calculate the distribution. By using multi-parton states 
factorizations of SSA with twist-3 operators have been studied in \cite{MS1,MS2,MSZ}. In \cite{MaWang} such multi-parton states 
have been employed to study the scale dependence of twist-3 operators relevant for SSA. 
\par 
We introduce the following state $\vert n(\lambda)\rangle$ as a superposition of single- and multi-parton states:
\begin{equation}
  \vert n (\lambda) \rangle =  \vert q \rangle + c_1 \vert q G \rangle  
  + c_2 \vert q GG \rangle + c_3 \vert q\bar q q \rangle + \cdots.
\label{MPS}   
\end{equation}
The state $\vert n\rangle$ is with the momentum $(P^+,0,0,0)$. All partons in the partonic states 
moving in the $z$-direction and the sum of their momenta is that of $\vert n \rangle$. The sum of  helicity of partons 
is $\lambda$. In the above we have suppressed quantum numbers of partons, which will be specified later.  
In principle one can introduce wave functions depending on momenta of partons. For simplicity, 
we take these wave-functions as $\delta$-functions and hence $c_i (i=1,2,3\cdots)$ are real constants. 
If we calculate $\tilde T_-$ of the state $\vert n(\lambda)\rangle$ 
instead of a hadron, we obtain nonzero contributions from interference between different partonic states. 
At tree-level, the contributions can be schematically written as:
\begin{eqnarray} 
\tilde T_- (x_1,x_2) &=& {\mathcal C}_1 \langle q(-) \vert {\mathcal O(x_1,x_2) } \vert q(+)g(-)\rangle
       + {\mathcal C}_2 \langle q(-) g(+) \vert {\mathcal O}(x_1,x_2)  \vert q(+) \rangle
\nonumber\\
      &&  + {\mathcal C}_3 \langle q(-) \bar q(-) \vert {\mathcal O} (x_1,x_2) \vert g(-)  \rangle
       + {\mathcal C}_4 \langle g(+) \vert {\mathcal O}(x_1,x_2) \vert q(+) \bar q(+) \rangle,  
\label{TTree}        
\end{eqnarray}
where $\pm$ in brackets indicate the helicity of partons. It should be noted that there are possible spectators. E.g., 
in the first term, the spectators must carry the total helicity $\lambda_s =0$ which can be a quark pair or gluon pair. 
In the second term the spectators must have $\lambda_s =-1$ because we have here $\lambda =-1/2$. For the last two terms 
we have $\lambda_s =\pm 1/2$, respectively. The contributions from spectators give overall factors as products of $\delta$-functions 
for each term. These overall factors are contained in the constants ${\mathcal C}_i$ which also depend on $c_i$ in Eq.(\ref{MPS}).  
Because of existence of spectator-partons, 
the parton states in the above are not necessarily with the total momentum $P$. 
\par 
Beyond tree-level, the four matrix elements of partons in Eq.(\ref{TTree}) receive corrections. 
These corrections make $\tilde T_-$ $\mu$-dependent.
We define the four matrix elements as: 
\begin{eqnarray} 
   T_{-qg} (x_1,x_2,y_0,z_0) &=& \langle q(p,-) \vert {\mathcal O}(x_1,x_2)  \vert q(p_1,+)g(p_2,-)\rangle,
\nonumber\\
   T_{+qg} (x_1,x_2,y_0,z_0) &=& \langle q(p_1,-)  g(p_2,+) \vert {\mathcal O}(x_1,x_2)  \vert q(p,+) \rangle,
\nonumber\\   
 T_{-q\bar q}  (x_1,x_2,y_0,z_0)  &=& \langle \bar q (p_2,-) q(p_1,-) \vert {\mathcal O}(x_1,x_2)  
    \vert g(p,-) \rangle,     
\nonumber\\
   T_{+ q\bar q}  (x_1,x_2, y_0,z_0)  &=& \langle g(p,+) \vert {\mathcal O}(x_1,x_2)  
    \vert  q(p_1,+) \bar q (p_2,+)  \rangle, 
\end{eqnarray} 
with the momenta:
\begin{eqnarray} 
  && p_1^+= y_0 P^+, \ \ \ \ p_2^+ =(z_0-y_0) P^+, \ \ \ \  p^+ = z_0 P^+.      
\end{eqnarray}
The color of the state $\vert qg\rangle $ is the same as the single-quark state. Details can be found in \cite{MS1,MSZ}. 
The four matrix elements are not independent. They are pair-wise related:
\begin{equation} 
T_{-qg} (x_1,x_2, y_0,z_0 )  =  T_{+q g}  (x_2, x_1, y_0, z_0 ) ,  \ \ \ \ 
T_{-q\bar q}  (x_1,x_2, y_0,z_0)  =   T_{+q\bar q}  (x_2,x_1,y_0,z_0). 
\label{SYM} 
\end{equation}
As we will see, the $\mu$-dependence of the four matrix elements determine the kernel ${\mathcal F}_-$ in different regions
of $\xi_{1,2}$. Because of this, the determined kernel will not depend on the states, i.e., the coefficients $c_i$ and ${\mathcal C}_i$. 
To determine the kernel in the full region of $\xi_{1,2}$, one also needs to calculate $\tilde T_-$ of the state $\vert \bar n \rangle$-- The charge-conjugated state of $\vert n\rangle$. However, $\tilde T_-$ of $\vert \bar n\rangle$ can be obtained from $\tilde T_-$ of $\vert n\rangle$ through charge conjugation. 
\par 

\begin{figure}[hbt]
\begin{center}
\includegraphics[width=13cm]{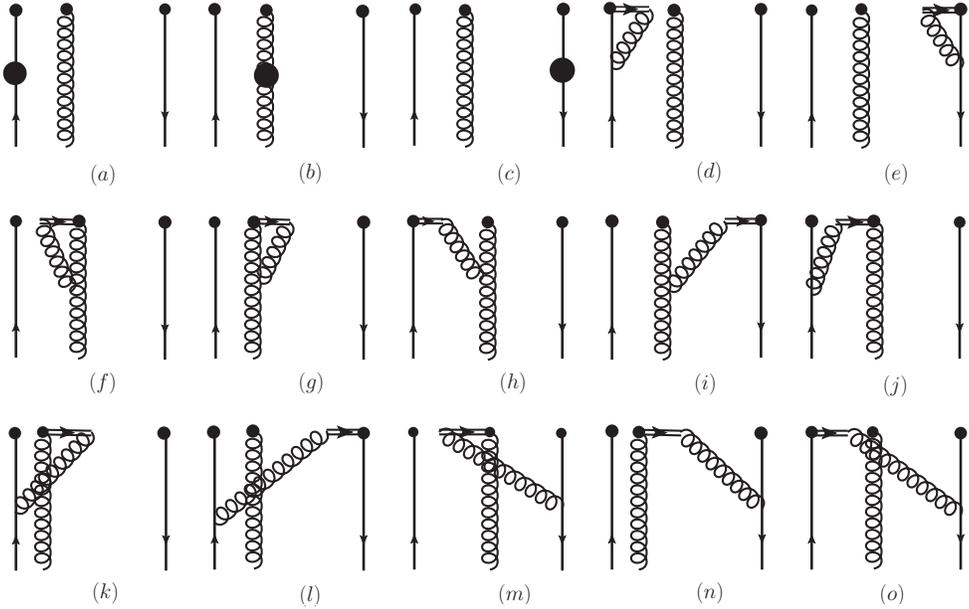}
\end{center}
\caption{A set of diagrams of one-loop corrections to the defined twist-3 matrix elements. This set
only contains the self-energy corrections and corrections from the gluon emission from a gauge link represented by 
a double line. }
\label{P2}
\end{figure}
\par
\par 
We take $T_{-qg}$ as an example to show how the corresponding contribution to ${\mathcal F}_-$ is determined. 
At tree-level, we have:
\begin{equation} 
 T^{(0)}_{-qg}  (x_1,x_2, y_0,z_0)  =  2 \pi g_s \sqrt{2 z_0 y_0 } (N_c^2-1) (z_0-y_0)  \delta (z_0-x_1) \delta (x_2-y_0) . 
\end{equation} 
It should be noted $z_0 > y_0$ because of $p_2^+ >0$. If we have the one-loop result $T^{(1)}_{-qg}$, we can substitute 
the results of $T^{(0,1)}_{-qg}$ into Eq.(\ref{GEV}) through Eq.(\ref{TTree}). Then we can find for $\xi_2>0$:
\begin{eqnarray} 
  g_s \alpha_s  {\mathcal F}_{-} (x_1,x_2,\xi_1,\xi_2) \theta (\xi_1-\xi_2)  =  \frac{1}{2(N_c^2-1) (\xi_1-\xi_2)\sqrt{2\xi_1\xi_2}} 
     \frac{\partial}{\partial \ln \mu }  T^{(1)}_{-qg}(x_1,x_2,\xi_2,\xi_1).     
\end{eqnarray} 
The $\theta$-function $\theta(\xi_1-\xi_2)$ appears because of $z_0-y_0 >0$. Therefore, ${\mathcal F}_-$ in the region of $\xi_1>\xi_2$ and $\xi_2>0$ 
is determined by $T^{(1)}_{-qg}$. Similarly, one can find that ${\mathcal F}_-$ in the region of $\xi_2 >\xi_1$ and $\xi_1>0$ 
is determined by $T^{(1)}_{+qg}$. Combining the two contributions one has the kernel in the region of positive $\xi_{1,2}$.   
\par 
\par 
\begin{figure}[hbt]
\begin{center}
\includegraphics[width=12cm]{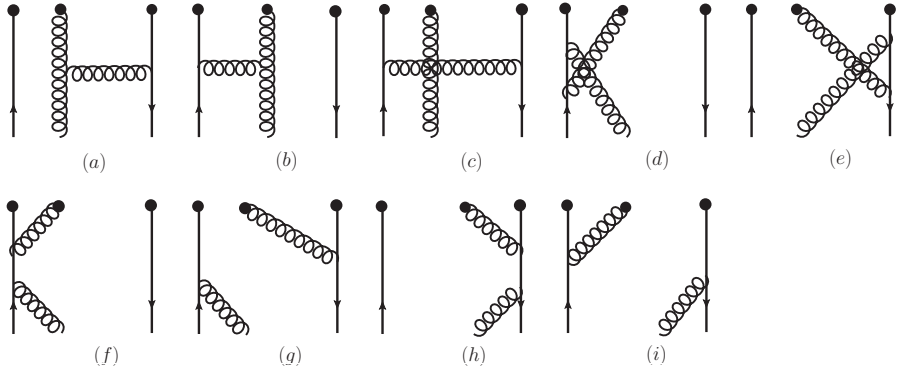}
\end{center}
\caption{Another set of diagrams for one-loop corrections of the defined twist-3 matrix elements.   }
\label{SFP-P6}
\end{figure}
\par
The one-loop correction to $T_{-qg}$ are from the diagrams given in Fig.1 and Fig.2. The calculation of these diagrams 
is rather standard. Therefore we give the result directly. We introduce the following function which is just the kernel 
in the region of $\xi_{1,2}>0$: 
\begin{eqnarray} 
    {\mathcal F}_{1} (x_1,x_2,\xi_1,\xi_2) &=&  \theta (x_2) \biggr  [ 
      \delta (\xi_1-x_1) f_1(\xi_2,x_1,x_2) + \delta (\xi_1-\xi_2-x_1+x_2)  f_0(\xi_2,x_1,x_2) 
\nonumber\\         
        &&  +\delta(\xi_2-x_2)  f_2(\xi_1,x_1,x_2) \biggr ] + (x_1 \leftrightarrow x_2,\xi_1 \leftrightarrow \xi_2 ),  
\end{eqnarray}
with 
\begin{eqnarray}
f_0 (\xi,x_1,x_2) &=& \frac{1}{2 N_c} \frac{x_2 \theta (\xi-x_2)}{\xi (x_2-\xi)_+}, 
\nonumber\\
f_1 (\xi,x_1,x_2) &=& \frac{N_c}{4} \biggr ( \theta (x_1-x_2) \theta(x_2-\xi) -\theta(x_2-x_1) \theta (\xi-x_2) \biggr ) 
           \biggr ( \frac{2}{(x_2-\xi)_+} -\frac{x_1-x_2}{(x_1-\xi)^2} -\frac{2}{x_1-\xi} \biggr )  
\nonumber\\
           && + \frac{N_c}{2} \theta(\xi-x_2)  \frac{x_2}{\xi}\biggr ( \frac{1}{(\xi-x_2)_+} -\frac{1}{\xi-x_1} \biggr ) 
             + \frac{1}{2}\delta (\xi-x_2) 
              \biggr [ \frac{3(N_c^2-1)}{4 N_c} 
\nonumber\\              
    &&  -\frac{N_c^2-1} {N_c} \ln x_2  
                 - N_c \ln \vert x_1-x_2 \vert \biggr ],                             
\nonumber\\    
f_2(\xi,x_1,x_2) &=& - \frac{N_c}{4(\xi-x_2)}   \theta (x_1) 
  \biggr [ \theta (x_1-\xi)\theta(x_2-x_1) \frac{x_1-x_2}{x_2 (\xi-x_2)} ( 3x_2 -2x_1-2\xi) 
\nonumber\\     
    && + 2 \theta (\xi-x_1) \theta (x_2-x_1) \frac{x_1}{\xi x_2} (x_2-x_1-\xi) 
\nonumber\\
   && - \theta (\xi-x_1) \theta (x_1-x_2) \frac{2x_2 x_1-3\xi x_2 - 2 x_1^2 + x_1 \xi  +2 \xi^2}{\xi(\xi-x_2)} \biggr ] 
\nonumber\\
  && +  \frac{N_c^2-1}{2N_c} \theta (x_2-x_1)\theta (x_1) \frac{x_1 (x_2-x_1)}{x_2^2 (\xi-x_2)}
\nonumber\\
  && + \frac{\theta(x_2-x_1) }{2 N_c (\xi-x_2)}    (x_2-x_1-\xi) 
\biggr [ -\frac{\xi-x_2 +x_1}{\xi(\xi-x_2)} \theta(-x_1) \theta (\xi-x_2+x_1)  
\nonumber\\
  && -  \theta (x_1) \left ( -\frac{x_1 \theta (x_2-\xi-x_1)}{x_2(\xi-x_2) } 
     + \frac{x_2-x_1}{\xi x_2} \theta (\xi-x_2 +x_1) \right ) \biggr ].     
\end{eqnarray} 
The $+$-distributions here are  defined as:
\begin{eqnarray} 
\int_0^{x} d z \frac{f(z)}{(x-z)_+} &=& \int_0^x d z \frac{f(z)-f(x)}{x-z } + f(x) \ln x ,  
\nonumber\\
\int_x^{1} d z \frac{f(z)}{(z-x )_+} &=& \int_x^1 d z \frac{f(z)-f(x)}{z-x} + f(x) \ln (1-x) .
\end{eqnarray} 
It should be noted that our $+$-distribution is not the standard $+$-distribution defined later. The function $f_0$ and $f_1$ 
are determined by the contributions from Fig.1. In general for each diagram with a gluon emitted from the gauge link 
there is a light-cone singularity. But, the sum is free from the singularity. 
\par 
Now we turn to the contributions from $T_{-q\bar q}$. At tree-level we have 
\begin{equation}
 T^{(0)}_{-q\bar q} (x_1,x_2, y_0, z_0)  = -2\pi g_s (N_c^2-1) \sqrt{2 y_0 (z_0- y_0 )}  z_0 \delta (x_1-y_0) 
\delta (x_2+z_0- y_0).  
\end{equation} 
With the one-loop result of $T_{-q\bar q}$ one can determine the kernel 
\begin{eqnarray} 
  g_s \alpha_s  {\mathcal F}_ {-}  (x_1,x_2,\xi_1,\xi_2) \theta (\xi_1)\theta(-\xi_2)  
    =  -\frac{1}{2(N_c^2-1) (\xi_1-\xi_2)\sqrt{-2\xi_1\xi_2} } 
     \frac{\partial}{\partial \ln \mu }  T^{(1)}_{-q\bar q} (x_1,x_2,\xi_1,\xi_1-\xi_2 ),  
\end{eqnarray}
The kernel determined from $T_{-q\bar q}$ is in the region of $\xi_1>0$ and $\xi_2<0$. From 
$T_{+q\bar q}$ the kernel in the region $\xi_1<0$ and $\xi_2>0$ can be obtained. 
The one-loop correction to $T_{-q\bar q}$ is still given by the diagrams in Fig.1 and Fig.2, where 
the quark line in the left side should be taken as an outgoing anti-quark. The calculation of these diagrams 
are again straightforward. The result can be summarized by the function
\begin{eqnarray} 
 {\mathcal F}_{2}  (x_1,x_2,\xi_1,\xi_2)&=& \delta(\xi_1-x_1)\theta (x_1) \biggr ( h_1(-\xi_2,x_1,x_2) 
       + h_2(-\xi_2,-x_2,-x_1) \biggr ) 
\nonumber\\       
     &&  + \delta(\xi_2-x_2) \theta (-x_2) \biggr ( h_1 (\xi_1,-x_2,-x_1) + h_2 (\xi_1,x_1,x_2) \biggr)
\nonumber\\
     &&   + \delta (\xi_1-\xi_2-x_1+x_2) \biggr ( \theta (-x_2) h_0 (-\xi_2,x_1,x_2) +  \theta (x_1)  h_0 (\xi_1,-x_2,-x_1) \biggr ),                   
\end{eqnarray} 
with the functions 
\begin{eqnarray} 
    h_0 (\xi,x_1,x_2)  &=& \frac{1}{2N_c} \theta (x_2+\xi) \frac{x_2}{\xi (\xi+x_2)_+}, 
\nonumber\\
    h_1(\xi,x_1,x_2) &=& \frac{N_c}{4} \biggr [   
    \theta (-x_2)  \theta (x_2+\xi)  \biggr ( \frac{\xi +2 x_1-x_2}{(\xi+x_1)^2} 
                 - \frac{ 2 x_2 }{\xi (\xi+x_1)} \biggr )\frac{x_1-x_2}{(\xi+x_2)_+}  
\nonumber\\
   && +\theta (x_1-x_2) \theta (x_2)   \frac{\xi +2 x_1-x_2}{(\xi+x_1)^2}\frac{x_1-x_2}{\xi+x_2} \biggr ] 
\nonumber\\   
     && + \frac{1}{2} \theta (-x_2) \delta (x_2+\xi) \biggr [ \frac{3}{4}\frac{N_c^2-1}{N_c} -\frac{N_c^2-1}{N_c}\ln \vert x_2\vert  -N_c \ln \vert x_1-x_2 \vert \biggr], 
\nonumber\\
 h_2(\xi,x_1,x_2) &=& -\frac{N_c}{4} \frac{1}{(\xi-x_2)^2} \biggr [ \theta (x_1) \theta(\xi-x_1) 
 \frac{ -2 \xi^2 -x_1\xi + 2 x_1^2 + 3 \xi x_2 -2 x_1 x_2}{\xi} 
\nonumber\\ 
    &&  + \theta (-x_1) \theta (x_1-x_2) \frac{x_1-x_2}{x_2} ( 2 \xi + 2x_1 - 3 x_2) \biggr ]
\nonumber\\
   &&  + \frac{1}{2 N_c} \frac{\xi+x_1-x_2}{(\xi-x_2)^2}\theta (-x_1) \theta (\xi-x_2+x_1) \biggr [ \theta (x_1-x_2) \frac{x_1}{x_2} 
         + \theta (x_2-x_1) \frac{\xi-x_2+x_1}{\xi} \biggr ]
\nonumber\\
   && +\frac{N_c^2-1}{2 N_c} \theta (-x_1) \theta (x_1-x_2) \frac{x_1(x_1-x_2)}{x_2^2(\xi-x_2)}.
\end{eqnarray} 
The function ${\mathcal F}_{2}$ is just the kernel in the region of $\xi_1>0$ and $\xi_2<0$. The 
kernel in the region of $\xi_1<0$ and $\xi_2>0$ can be obtained from $T_{+q\bar q}$ through the relation in
Eq.(\ref{SYM}). 
\par 
Finally, we have the kernel in the full region of $\xi_{1,2}$ as:
\begin{eqnarray} 
{\mathcal F}_-(x_1,x_2,\xi_1,\xi_2) &=& \theta(\xi_1)\theta(-\xi_2)  {\mathcal F}_{2} (x_1,x_2,\xi_1,\xi_2)
     +\theta(\xi_2)\theta(-\xi_1)  {\mathcal F}_{2} (x_2,x_1,\xi_2,\xi_1)
    \nonumber\\
  && + \theta (\xi_1)\theta(\xi_2) {\mathcal F}_{1} (x_1,x_2,\xi_1,\xi_2) 
   + \theta (-\xi_1) \theta(-\xi_2) {\mathcal F}_{1} (-x_2, -x_1, -\xi_2, -\xi_1). 
\label{main}    
\end{eqnarray}
The kernel in the region with $\xi_{1,2}<0$ can be obtained from that in the region with $\xi_{1,2}>0$ 
through charge-conjugation. Eq.(\ref{main}) is our main result.   
Since $\tilde T_F$ and $\tilde T_{\Delta}$ are not mixed under renormalization, the correct kernel should satifiy:
\begin{equation} 
 {\mathcal F}_{- } (x_1,x_2,\xi_1,\xi_2)  =  {\mathcal F}_{- } (x_2,x_1,\xi_2,\xi_1). 
\end{equation}
Our result has the property. Therefore, taking the part of ${\mathcal F}_-$ symmetric between $x_1$ and $x_2$, one can obtain 
the evolution of $\tilde T_F$ as a convolution with $\tilde T_F$. The anti-symmetric part of ${\mathcal F}_-$ gives the kernel 
of the evolution of $\tilde T_{\Delta}$.  
\par
The evolution kernel for arbitrary $x_{1,2}$ is rather lengthy. But the convolution in Eq.(\ref{GEV}) is a one-dimensional integral 
at one-loop. In some special cases, the kernel is simplified. E.g., for $x_1=x_2= x >0$, i.e., the case where the gluon carries zero 
momentum fraction entering a hard scattering,  we have:
\begin{eqnarray} 
\frac{\partial}{\partial \ln \mu} \tilde T_{ F }(x,x,\mu)    
     & =&  \frac{\alpha_s}{\pi} \biggr \{ -\frac{N_c^2 +3}{4 N_c} \tilde T_F (x,x,\mu) 
         +  \int_x^1 \frac{dz}{z} \frac{1}{(1-z)_+} \left ( N_c\tilde T_F(x,\xi) -\frac{z}{N_c} \tilde T_F (\xi,\xi)\right ) 
\nonumber\\
        && \ \ \ \  + \frac{1}{N_c} \int_0 ^{1-x} d\xi \frac{\xi}{(\xi+x)^2} \tilde T_F (x,-\xi) \biggr \}, 
\nonumber\\ 
  \frac{\partial}{\partial \ln \mu} \tilde T_{ \Delta  }(x,x,\mu)    
     & =& 0
\label{SG}      
\end{eqnarray}       
with $z =x/\xi$. Here the $+$-distribution is the standard one defined as:
\begin{equation} 
  \int_0^1 d z \frac{\theta (z-x)}{(1-z)_+} t(z) = \int_x^1 d z \frac{t(z)-t(1)}{1-z} + t(1) \ln (1-x). 
\end{equation} 
The result in Eq.(\ref{SG}) agrees with that in \cite{KQ2}. There are also cases in which a quark carries zero momentum fraction 
entering a hard scattering. In these cases, one has either $x_1=0$ or $x_2 =0$. The evolutions in these cases are:          
\begin{eqnarray}        
   \frac{\partial}{\partial \ln \mu} \tilde T_{ - }(0,x,\mu)   
   &=&  \frac{\alpha_s}{\pi} \biggr \{ \int_x^1 \frac{d z}{z } \biggr [  \left ( -\frac{N_c}{2}( 1+ z+ z^2) \tilde T_-(0,\xi) 
     + \frac{1}{2N_c} \tilde T_-(\xi,x) +\frac{1}{2N_c} 
     \tilde T_- (\xi-x,\xi) \right ) 
\nonumber\\      
      && +  \frac{1}{(1-z)_+ } \left ( N_c\tilde T_-(0,\xi) +\frac{N_c}{2} \tilde T_-(x-\xi,x) -\frac{1}{2 N_c} \tilde T_-(\xi-x,\xi)  \right ) 
\nonumber\\      
  &&  + \frac{(1-z)^2}{2 N_c} \tilde T_-(0,-\xi) 
   + \frac{3(N_c^2-1)}{4 N_c} \tilde T_-(0,x)  \biggr \}.                      
\end{eqnarray} 
Changing every $\tilde T_-(x_1,x_2)$ to $\tilde T_-(x_2,x_1)$ in the above equation, we obtain the evolution 
of $\tilde T_-(x,0)$.             
\par
The two studied twist-3 distributions do not mix in the evolution with other twist-3 distributions mentioned in the beginning, e.g.,
with $e(x)$ and $h_L(x)$. But the evolution of the distribution $e(x)$ or $h_L(x)$ does mix with $\tilde T_F$ and 
$\tilde T_\Delta$\cite{BBKT,KoNi,BD}. One may derive the evolution of $e(x)$ and $h_L(x)$ through their relations
to $\tilde T_F$ and $\tilde T_\Delta$, respectively. But the derivation is very tedious. It has been shown in \cite{BBKT} that in the large-$N_c$ limit
the evolution of $e(x)$ or $h_L(x)$ is simplified and obeys DGLAP-type equations. The result for this in momentum fraction space 
can be found in \cite{BD}. These evolutions in the large-$N_c$ limit can also be derived with the approach used here. We take 
$e(x)$ as an example to give our result. $e(x)$ is defined in the light-cone gauge as:
\begin{eqnarray}
  e(x) = P^+ \int\frac{d y}{4\pi} e^{-iy x P^+ } 
   \langle P \vert \bar \psi ( y n)   \psi (0) \vert P\rangle. 
\end{eqnarray}    
To derive the evolution of the two-parton twist-3 distribution  in the large-$N_c$ limit one can use the same approach for deriving the evolution of three-parton twist-3 distributions. The derivation is straightforward. We have for $x>0$: 
\begin{equation}
\frac{\partial e(x)}{\partial \ln\mu} = \frac{\alpha_s N_c}{2 \pi} 
\biggr [  \frac{1}{2} e (x) +  \int_x ^1 \frac{dz}{z}  e (\xi) \left ( \frac{ 2 }{(1-z)_+ } + 1 \right )  \biggr ] + {\mathcal O}(N_c^{-1}), 
\end{equation}   
with $z=x/\xi$. This result agrees with that in \cite{BD}. Although the evolution of $e(x)$ and $h_L(x)$ gets simplified 
in the large-$N_c$ limit, the evolutions of $\tilde T_{F,\Delta}$ are not much simplified in the limit. This can be 
observed from our results. This supports the observation made in \cite{JiOs} that simplifications of evolutions of higher-twist 
distributions can be accidental.
\par
To summarize: We have calculated the twist-3 distributions of multi-parton states. The distributions are defined 
with chirality-odd quark-gluon operators. The evolution kernel of the distributions has been obtained by the calculation 
at one-loop.  The evolution at one-loop is an one-dimensional convolution. In some special cases the evolution 
takes a short form. We have also derived the evolution of $e(x)$ in the large-$N_c$ limit and found an agreement with existing 
results.

\par\vskip20pt
\noindent
{\bf Acknowledgments}
\par
The work of J.P. Ma is supported by National Nature
Science Foundation of P.R. China(No. 10975169, 11021092, 11275244), and the work of Q. Wang is supported by National Nature
Science Foundation of P.R. China(No.10805028).     
\par

\end{document}